\documentclass[a4paper,11pt,showkeys,nofootinbib,longbibliography,maps,superscriptaddress]{revtex4-1}
\usepackage{hyperref}  
\usepackage{amsmath, bm}  
\usepackage{graphicx}
\usepackage{url}
\usepackage{soul}
\usepackage{color}
\usepackage[percent]{overpic}

\begin{document}

\title{Direct Measurement of Individual Dose from External Exposure by Glass-badge among All Citizens of Date City 5 to 51 Months after the Fukushima NPP Accident (series):\\
2: Prediction of Lifetime Additional Effective Dose and Evaluating the Effect of Decontamination on Individual Dose}

\author{Makoto Miyazaki}
\affiliation{Department of Radiation Health Management, Fukushima Medical University, Fukushima 960-1295, Japan
}
\author{Ryugo Hayano}
\affiliation{Department of Physics, The University of Tokyo, Tokyo 113-0033, Japan
}

\begin{abstract}

In the first paper of this series, we  showed that the ratio $c$ of individual dose to ambient dose did not change with time in Date City, Fukushima Prefecture, after the Fukushima Daiichi Nuclear Power Plant accident. The purpose of the present paper, the second in a series, is to estimate the lifetime doses of the Date City residents, based on continuous glass badge monitoring data, extrapolated by means of the ambient-dose-rate reduction function obtained from the airborne monitoring data.
As a result, we found that the  external exposure contribution to the mean additional lifetime dose of residents living in Date City  is not expected to  exceed 18 mSv.
In addition, effects of decontamination on the reduction of individual doses were not evident.
 This method of combining  individual doses and the ambient doses, as developed in this study, has made it possible to predict with reasonable certainty the lifetime doses of residents who continue to live in this radiologically contaminated area.
\end{abstract}
\keywords{Fukushima Dai-ichi Nuclear Power Plant accident, glass-badge personal dosimetry, estimation of lifetime doses, effects of decontamination on individual doses}
\maketitle

\section{ INTRODUCTION}

In the previous paper in this series~\cite{miyazaki2016}, we analyzed the  
results obtained from the 
``glass badge'' individual dosimeters (radio-photoluminescence (RPL) glass dosimeters: Glass Badge\textsuperscript\textregistered) 
used by residents of Date City, Fukushima Prefecture, from 5 to 51 months after the Fukushima Daiichi Nuclear Power Plant (FDNPP) accident, and concluded as follows:
\begin{itemize}
\item[1)] The individual doses obtained by individual dose monitoring with glass badges and the ambient doses at the residences of the glass-badge participants (``grid dose'' as defined in Ref.~\cite{miyazaki2016}) estimated from the airborne monitoring  survey carried out in the same period~\cite{sanada} were proportional with a coefficient of 0.15. 

\item[2)] The obtained coefficient of 0.15 did not change with time, across the six airborne surveys conducted between November 2011 (4th survey) and November 2014 (9th survey). 

\item[3)] As a result, the conversion factor from the ambient dose rate to the personal dose, 0.6, adopted by the Ministry of the Environment~\cite{safetyassessment}, was found to be about four times larger than the coefficient 0.15 obtained by these actual measurements.
\end{itemize}
In particular, conclusion 2) implies that the individual doses and the ambient doses decreased at the same rate. This further suggests that the lifetime doses of the residents who continue to live in the contaminated zones can be inferred from ambient doses obtained from the airborne monitoring surveys.

In this paper, the second in a series of three papers based on the Date City glass badge data, we first obtain a function $f(t)$ for the temporal change of the ambient dose rate based on the airborne monitoring data. We then compare the time-integrated ambient-dose-rate curve $F(t)=\int^t f(\tau) d\tau$ to the additional cumulative doses measured from 5  to 51 months after the FDNPP accident.  We further provide the estimates for the lifetime cumulative doses of the residents, by making use of $F(t)$.

In addition, we examine the effect of decontamination on the individual doses of the residents, who continued to use  glass badges and lived in the designated decontamination area throughout the study period.  We then discuss the implications of our findings regarding the effectiveness of decontamination in the reduction of long-term cumulative doses.

This paper discusses the results of these two studies and aims to establish a method to reliably predict the long-term cumulative external doses of residents living in the contaminated areas.

\section*{ETHICS STATEMENT}

This study was approved by the ethics committee of the Fukushima Medical University (Approval No. 2603).

\section{MATERIALS AND METODS}

\subsection{Modeling the temporal change of the ambient dose rate:}

The Japanese government regularly monitors the area radiologically affected by the Fukushima Daiichi NPP accident using  aircraft flying at an altitude of about 300\,m~\cite{sanada} and converts the measured values to the ambient dose rate $\dot H^{*}_{10}$ at 1\,m above the ground. The averaged data within the 250\,m grids have been disseminated as maps and numerical data~\cite{map}. The aerial survey was first conducted in April 2011 and the most recent was the tenth, conducted in November 2015. In our previous paper~\cite{miyazaki2016}, we used data from the fourth through ninth airborne monitoring surveys, which overlapped in timing with the glass-badge measurements from Sept. 2011 to 2014Q3. Note that when obtaining the additional exposure dose rate from the ambient dose rate obtained by airborne monitoring, 0.04\,$\mu$Sv h$^{-1}$, the value corresponding to the natural-radiation dose rate adopted by the Ministry of the Environment~\cite{safetyassessment},   is subtracted.

\begin{figure}[bth]
\includegraphics[width=0.5\columnwidth]{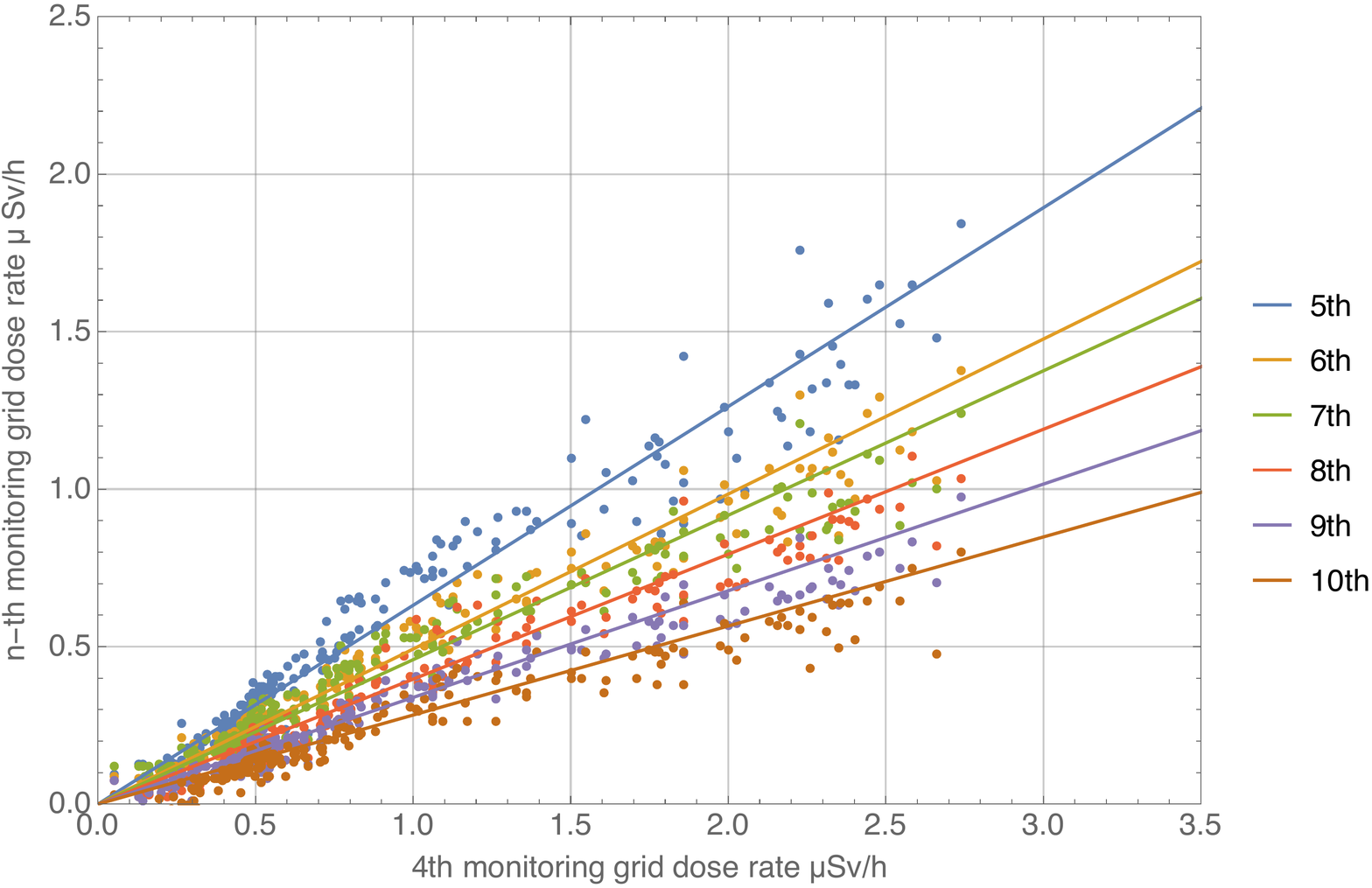}
\caption{\label{fig1l} \footnotesize 
The ambient-dose rates of 5th (blue) to 10th (brown) airborne monitoring surveys (ordinate) are plotted for each grid point in Date City, verses those of the fourth monitoring survey (abscissa).
 The six straight lines fitted through each of the 5th to 10th survey points are drawn in the corresponding colors.\\}

\includegraphics[width=0.5\columnwidth]{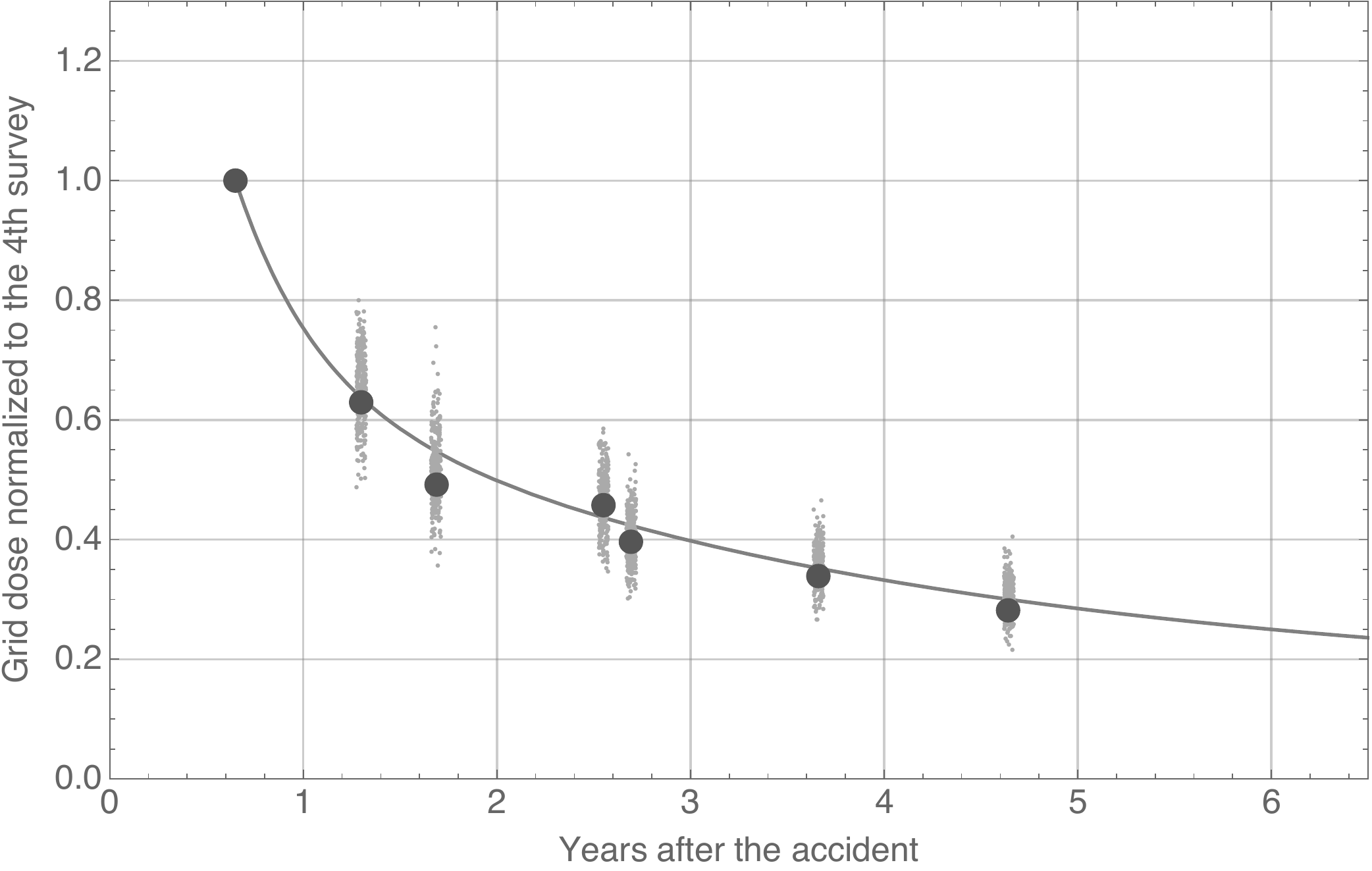}
\caption{\label{fig1r}\footnotesize The ratios of the $n$th ($n = 5$ to 10) grid dose to the 4th grid dose are plotted as a function of elapsed time since the accident. In order to help alleviate overlaps of points, the data were horizontally  jittered, and the average ratios are shown in filled circles.
 The curve is a fit to Eq.~(1) through all the data points, the best fit parameters being $a_{\rm fast} = 0.68 \pm 0.01$ and $T_{\rm fast} = 0.32 \pm 0.01$ y.}
\end{figure}

In order to extract the temporal change of the ambient dose rate in Date City, we first tabulated, for each grid point in Date City and for the fifth to tenth surveys, the background-subtracted ambient-dose-rate ratio derived from each monitoring survey   subsequent to (but not including)   the fourth.  The results are shown in Fig.~\ref{fig1l}. As shown, there is a good linearity between the ambient dose rates of $n$th and 4th surveys. We therefore fitted a straight line (also shown in Fig.~\ref{fig1l}) for each survey from the 5th to the 10th. The slope of each straight line represents the average reduction of the ambient dose rate relative to the 4th survey, which we plot with filled circles in Fig.~\ref{fig1r}, versus years after the accident.

Next, the time dependence of the ambient dose rates was modeled by the following function, known from previous studies~\cite{kinase,ecological}, which contains the physical half-lives of Cs-134 and Cs-137 ($T_{134}=2.06\rm y$ and $T_{137} = 30.17\rm y$), fast- and slow-ecological half-lives ($T_{\rm fast}$ and $T_{\rm slow}$) and the fraction of the fast-ecological decay component ($a_{\rm fast}$),
\begin{equation}
f(t)=\frac{\dot{H}^*_{10}(t)}{\dot{H}^*_{10}(0.65)}=\left\{a_{\rm fast} 2^{-t/T_{\rm fast}}+ (1-a_{\rm fast})2^{-t/T_{\rm slow}}\right\}\cdot \frac{(k \times 2^{-t/T_{134}}+2^{-t/T_{137}})}{k+1},
\end{equation}
where ${\dot{H}^*_{10}(0.65)}$ denotes that this function is normalized to the ambient dose rates of the 4th airborne survey conducted at $t=0.65 \rm y$.
Here, the coefficient $k$ was derived as follows:
Radioactive contamination in Date City is due primarily to radioactive materials released from the FDNPP No.\,2 reactor on March 15, 2011~\cite{chino}. Analysis of radioactivity in each nuclear reactor of FDNPP released from JAEA in 2012~\cite{jaea2012} at the shut down point of the second reactor is: $2.76 \times 10^8$\,Bq ($^{134}$Cs) and $2.55 \times 10^8$\,Bq ($^{137}$Cs). Together with the kerma-rate conversion coefficients 4.68 ($^{134}$Cs) and 1.72 ($^{137}$Cs) nGy h$^{-1}$ per kBq m$^{-2}$~\cite{icrp74}, the coefficient of $k=2.95$ was derived.

\subsection{Estimated Additional Lifetime Doses:}

Date City started measuring individual doses with glass badges in August 2011, when measurements were done for pregnant women and children aged 15 years and younger for one month. From September 2011, the measurement periods were extended to 3 months (1 Quarter) and measurements are still being continued at the time of writing (Table 1 of Ref.~\cite{miyazaki2016}). The dosimeter supplier, Chiyoda Technol Corporation, has been using the value of 0.54\,mSv\,y$^{-1}$ as the contribution of the natural background radiation throughout Fukushima Prefecture since the FDNPP accident. This was measured in Oarai in Ibaraki Prefecture, and is subtracted from the glass-badge readings when they report the additional external doses to the relevant municipalities~\cite{nomura}, including Date City.

In the present paper, we selected the subjects who  held the glass badges continuously from 2011Q3 to 2015Q1 (from September 2011 to June 2015), as this makes it possible to directly obtain their cumulative individual doses  for the said period. The numbers of subjects thus selected were $n = 476$ for zone A, $n = 693$ for zone B and $n = 3280$ for zone C (see Sec.\,\ref{zoning} for the definition of the three zones). The additional cumulative external dose for the first 4 months after the accident, before the glass-badge measurement period, was assumed to be 1.4\,mSv, which is the average value of the external  dose estimated for the residents living in the northern Fukushima Prefecture  region, as was published in the basic survey included in the Fukushima Health Management Survey~\cite{healthsurvey}.

The cumulative individual doses $H_P (t)$ were then compared to the following function,
\begin{equation}
H_{P}^i (t) = \int_{t_1}^t \left(\dot H_{10}^{*\, i}(0.65)\times  c^i \times  f(\tau) \right) d\tau +I,
\end{equation}
where $f(\tau)$ is the ambient dose reduction function obtained in Eq\,(1),  $i$ is for zone A, B and C. The mean ambient dose rates at the time of the 4th airborne monitoring were $\dot H_{10}^{*\, A}(0.65)=2.1 \mu\rm Sv\,h^{-1}$, $\dot H_{10}^{*\, B}(0.65)=1.4 \mu\rm Sv\,h^{-1}$ and $\dot H_{10}^{*\, C}(0.65)=0.8 \mu\rm Sv\,h^{-1}$, $c^i$ is the conversion coefficient  for zone $i$ (see below), $t_1=0.39$ is the beginning of the glass-badge survey (August 1, 2011), and $I$ is the estimated average external dose during the first 4 months of the accident (see above). 

The conversion coefficient $c$,
\begin{eqnarray}
\left<c\right>\equiv \left<\frac{\mbox{individual  dose rate}}{\mbox{grid dose rate}}\right>, \label{eq:correlation}
\end{eqnarray}
calculated in the previous paper, $c=0.15\pm 0.03$~\cite{miyazaki2016}, was the average of all the glass-badge survey participants.  For the present study, we recalculated $c$ for each zone, for those who continuously held the badge. The results are:
$c^{\rm A} = 0.10$, $c^{\rm B} = 0.12$, and $c^{\rm C} = 0.15$.

Finally, the function $H_P^i(t)$ was extrapolated to $t=70\,\rm y$ to estimate the average  additional lifetime external dose.

\subsection{Difference in the decontamination method for each zone in Date City:\label{zoning}}

Date City pioneered the decontamination efforts, such as top soil removal, as described in the decontamination implementation plan (1st edition) published in October 2011~\cite{datedecontamination1}.
In the 2nd edition, published in August 2012~\cite{datedecontamination},  the city was divided into three zones, Zone A ($>3.5 \mu \rm Sv\, h^{-1}$), Zone B (between $1 \mu\rm Sv\, h^{-1}$ and $3.5\mu\rm Sv \,h^{-1}$), and Zone C ($<1\mu \rm Sv\,  h^{-1}$). Zoning was set mainly for the purpose of determining the methods and priority of decontamination, and to carry out decontamination properly and promptly. According to the implementation plan, the residential premises and the surrounding forest up to 20\,m from  residential boundaries were to be decontaminated in zone A, residential premises were to be decontaminated in zone B, and ``hot spots'' were to be cleaned in zone C. The method  for Zone A is equivalent to the method described in the Decontamination Guidelines, published  by the Ministry of the Environment~\cite{moeplan}. The decontamination of  residential premises based on the  implementation plan in Date City was completed in March 2014.

\subsection{Evaluation of the influence of decontamination on individual doses:}

\begin{figure}[bt]
\includegraphics[width=0.3\columnwidth]{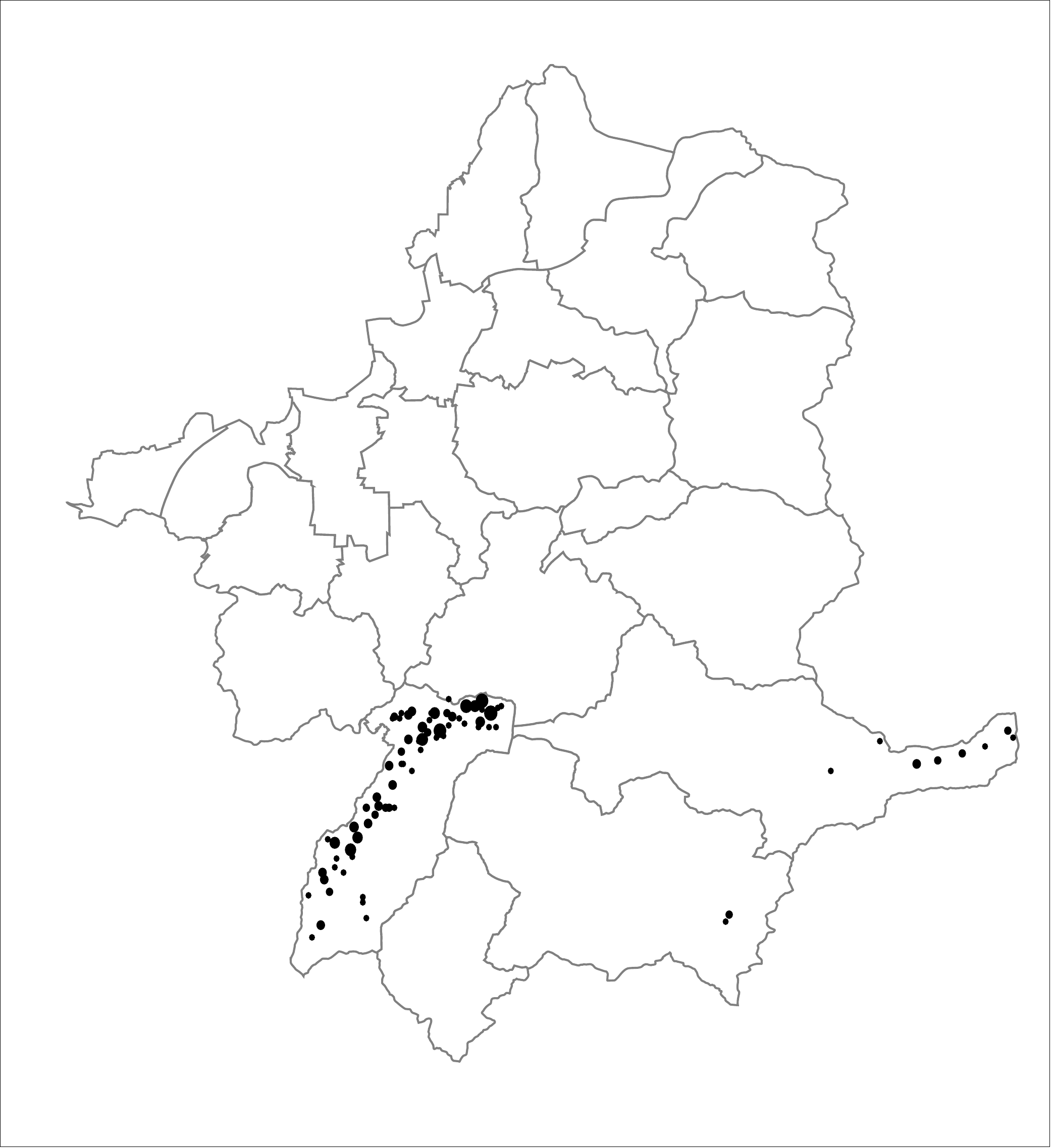}
\hfill
\caption{\label{fig2l} \footnotesize The geographical distribution of 132 houses in Zone A of Date City, for which the decontamination work took place in Q3 of  2013(October to December, 2013). }
\end{figure}

\begin{figure}[h]
\includegraphics[width=.5\columnwidth]{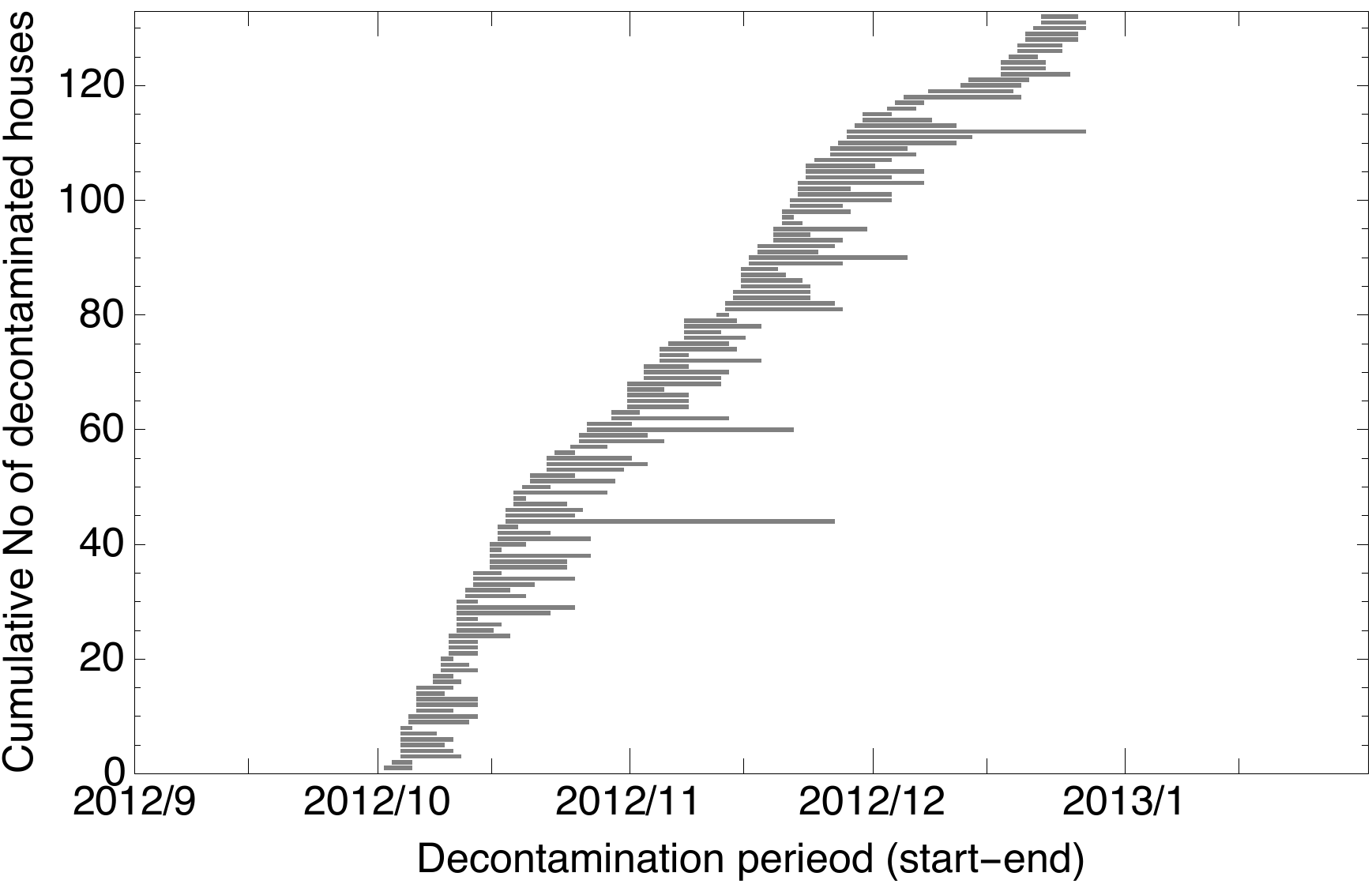}
\caption{\label{fig2r} \footnotesize The start and end of the decontamination work of the 132 houses are indicated by horizontal bars. }
\end{figure}

Since the decontamination methods adopted by Date City differed for each zone,  in order to correctly evaluate the effect of decontamination on individual doses it is necessary to select a single zone (i.e., common decontamination method), and to select the subjects who continuously  held the glass badges while their houses were decontaminated. We focused on 1,700 houses where decontamination was done in zone A, among which all family members continuously  held glass badges from 2011Q3 to 2014Q1 (September 2011 to June 2014), and  decontamination was carried out during 2012Q3 (from October to December 2012). This narrowed  the number of subjects to  425,  living in 132 houses. The glass badge data and the GIS information of the subjects were anonymized by Date City, and was provided to the  authors. Figures~\ref{fig2l} and \ref{fig2r} illustrate the approximate positions of the 132 houses targeted and the time periods when decontamination was performed respectively. The distribution of the measured values of the glass badges of the selected subjects over time was evaluated, as was the influence of decontamination on the glass badge measurement values.

\hfill

\section{RESULTS}

\subsection{Temporal change of the ambient dose rate:}

By fitting the  model Eq.\,(1)  to the ambient dose rate ratios of the 5th to 10th airborne surveys  relative to the 4th survey (filled circles in Fig.~\ref{fig1r}), the parameters in Eq.\,(1) have been determined as follows: $a_{\rm fast} = 0.68\pm 0.01$, $T_{\rm fast} = 0.32\pm0.01$\,y, $T_{\rm slow} > 700$\,y (therefore, the slow ecological component can be ignored for our purposes). 
The line in Fig.~\ref{fig1r} is the best-fit curve.

\subsection{Estimation of additional lifetime doses:}

\begin{figure}
\begin{center}

\begin{overpic}[width=0.55\textwidth]{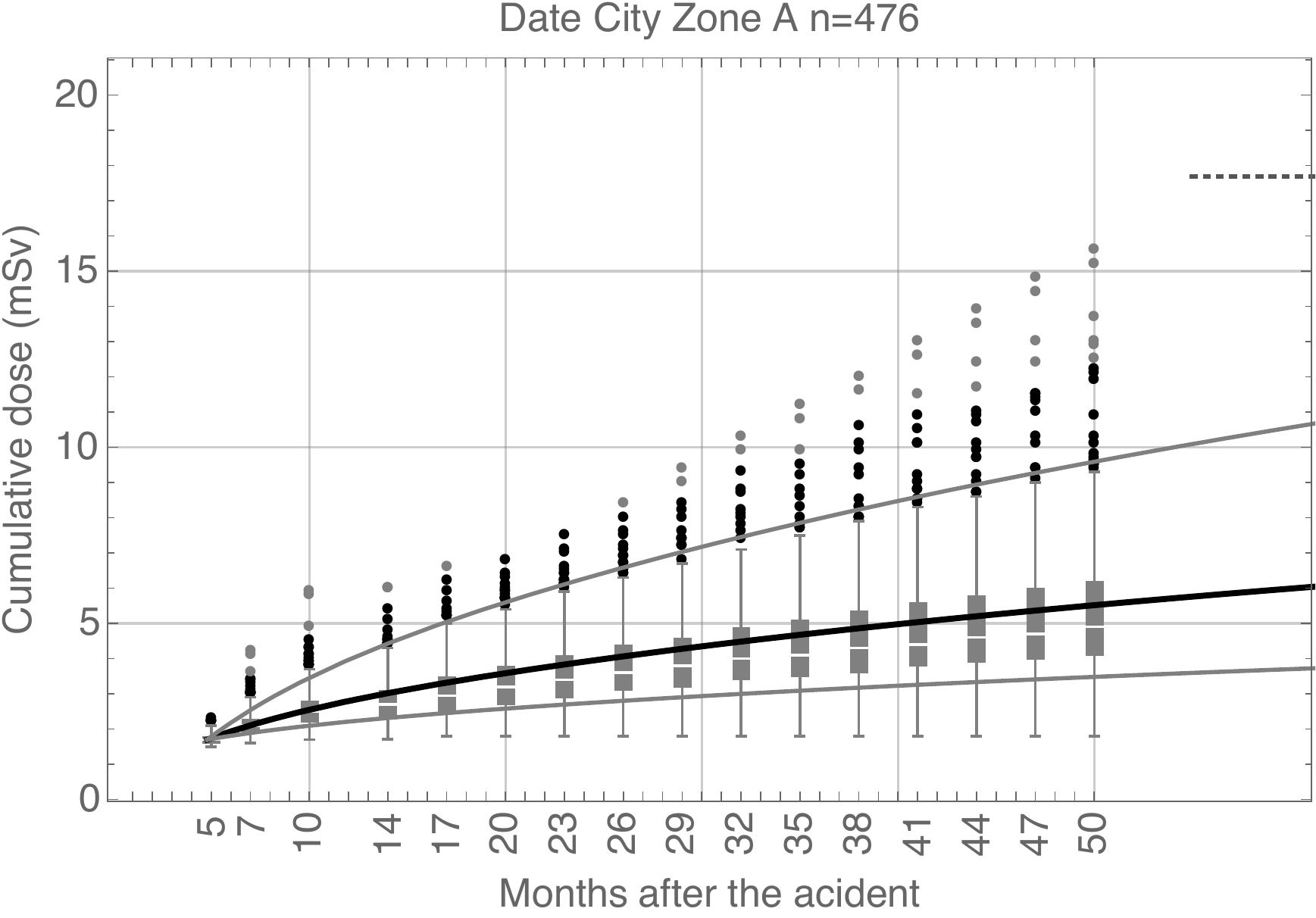}
\put(10,60){\large\bf 1)}
\end{overpic}

\vspace*{0.5cm}

\begin{overpic}[width=0.55\textwidth]{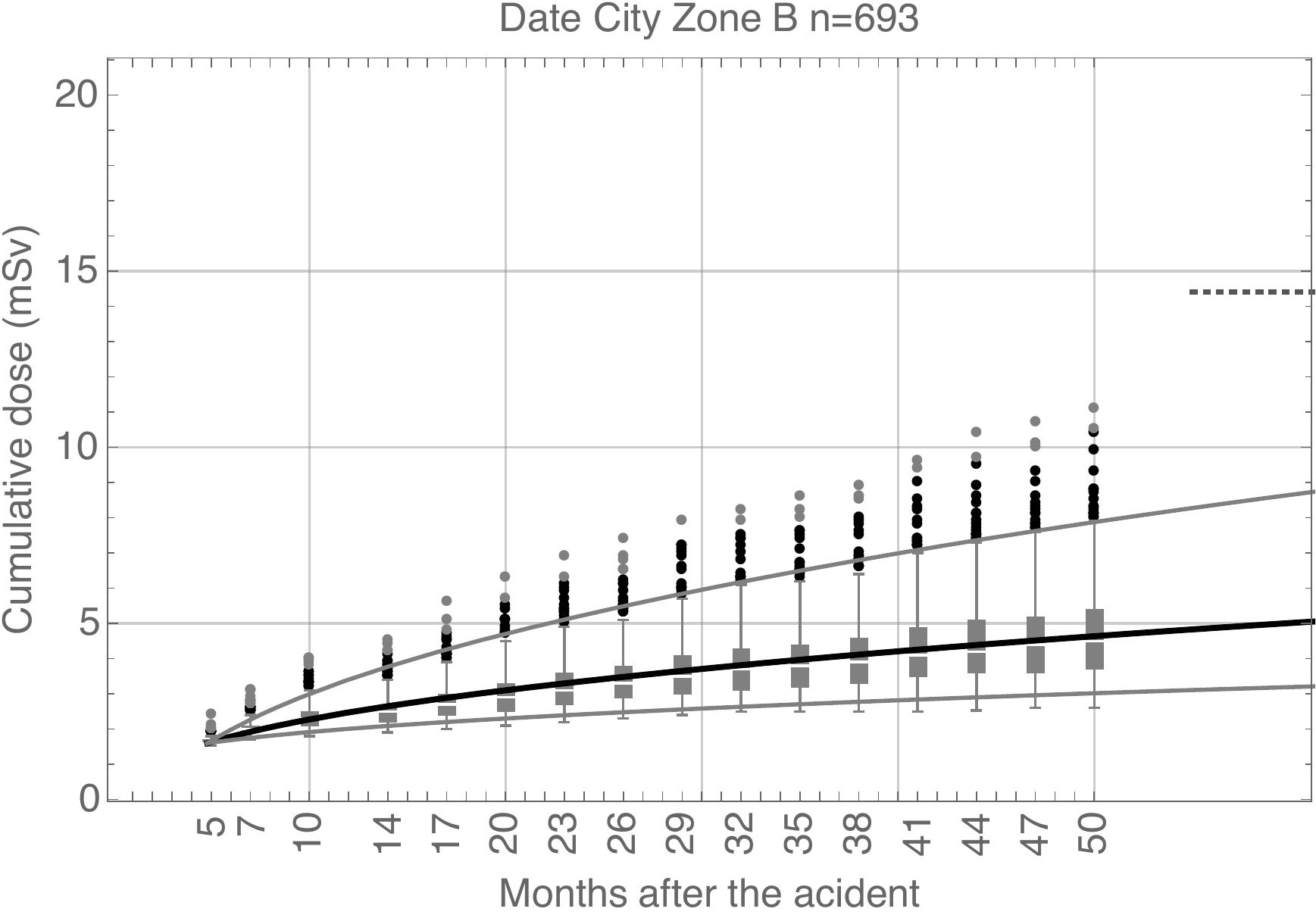}
\put(10,60){\large\bf 2)}
\end{overpic}

\vspace*{0.5cm}

\begin{overpic}[width=0.55\textwidth]{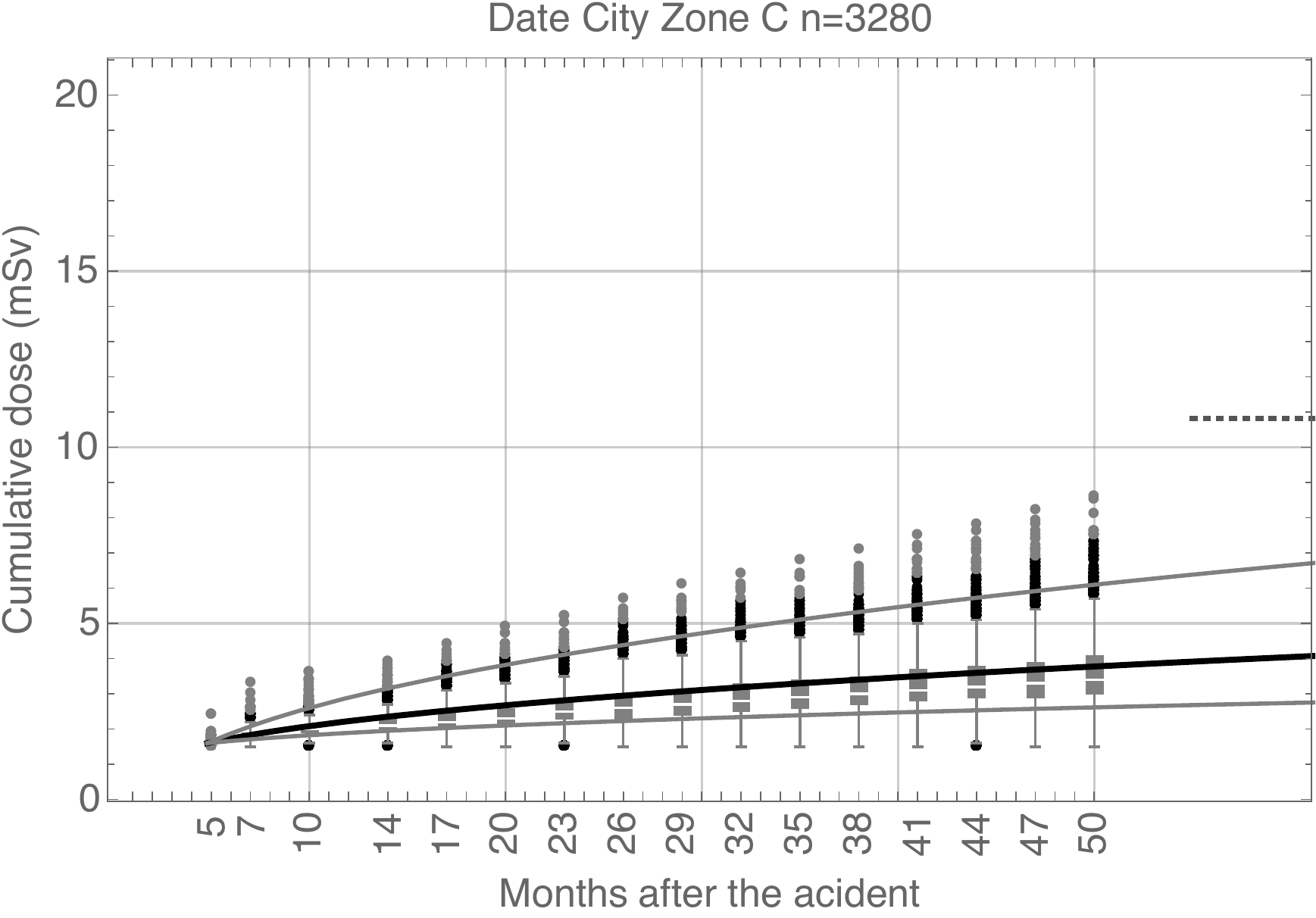}
\put(10,60){\large\bf 3)}
\end{overpic}

\end{center}
\caption{\label{fig5}\footnotesize Box-and-whisker plots of the cumulative individual doses of Date City residents (Fig.~\ref{fig5}-1) is for Zone A, 5-2 is for Zone B and 5-3 is for Zone C), who continuously held glass badges during the study period.  The boxes cover 25-percentile to 75-percentile of the distribution, and the whiskers cover the 1-percentile to 99-percentile of the distribution. The dots represent outliers. The dark solid curve is the estimated median $H_p (t)$ (Eq.~(2)), while upper and lower light curves correspond to the 1- and 99-percentile estimates, respectively. The dotted line along the right vertical axis of the graph represents the estimated median lifetime doses (up to 70 years) from external exposure pathways, 18 mSv for zone A, 15 mSv for zone B, and 11 mSv for zone C.}
\end{figure}

Figures 5-1, 5-2, and 5-3, show the box-and-whisker plots for the accumulated individual doses for the Date City residents in Zone A, B and C, respectively, who continuously held a glass badge. The closed circles are outliers ($>99$-percentile).

For each plotted cumulative-dose distribution, the integral curve of the individual dose estimated from the grid dose rates for each zone $ H_P^i (t), (i=A, B, C)$ is superimposed (dark solid line). Also shown in light solid lines are the 1-percentile and 99-percentile curves estimated respectively by taking the 1-percentile and 99-percentile values of the coefficient $c^i$ in Eq.\,(2).
The integral curve of the measured individual dose and the individual dose estimated from the grid dose show fair agreements with the 50-percentile and 99-percentile curves in all zones.

By extrapolating $H_P^i(t)$ to $t=70\rm y$ (and by adding the estimated external dose for the initial 4 months) we estimated the additional lifetime dose for each zone.  The median of the cumulative dose (and the 99-percentile value) were presumed to be 18 mSv (35 mSv) for Zone A, 15 mSv (28 mSv) for Zone B, and 11 mSv (20 mSv) for Zone C, respectively. In each of Figs.\,5-1, 2 and 3, the median of the cumulative dose is indicated by a dotted line.

\subsection{Evaluation of the influence of decontamination on individual dose:}

A box-and-whisker plot of individual doses (converted to the hourly dose $\mu$Sv h$^{-1}$) of the 425 Zone-A residents, from 2011Q3 to 2014Q1, is shown in Fig.\,6. 

The best-fit  curve $f(t)$, multiplied by the mean ambient dose rate at the time of the 4th airborne monitoring ($\dot H^{*\,A}_{10} (0.65)=2.1 \mu\rm Sv h^{-1}$) and the ambient-dose-to-individual-dose coefficient for Zone A ($c^A =0.1$), is superimposed on the plot, and the timing of the decontamination (2012Q3) is indicated by an arrow.  In Fig.~7, $H_P^A(t)$ is superimposed on the distribution of cumulative individual doses presented in Fig. 6.  As shown, no clearly-observed effect of decontamination on the cumulative dose was evident.
The median of individual 3-months doses for these subjects decreased by 62\%, from 0.8 mSv (2011Q3) to 0.3 mSv (2013Q3), in about two years. Correspondingly, the ambient dose rate shown by the seventh airborne monitoring survey (October 2013)  in Date City decreased by about 60\% compared with the value of the 4th (November 2011), as shown in Fig.~\ref{fig1r}. Thus, the reduction of the individual doses and the airborne monitoring doses remained almost unchanged.

\begin{figure}[h]
\includegraphics[width=.5\textwidth]{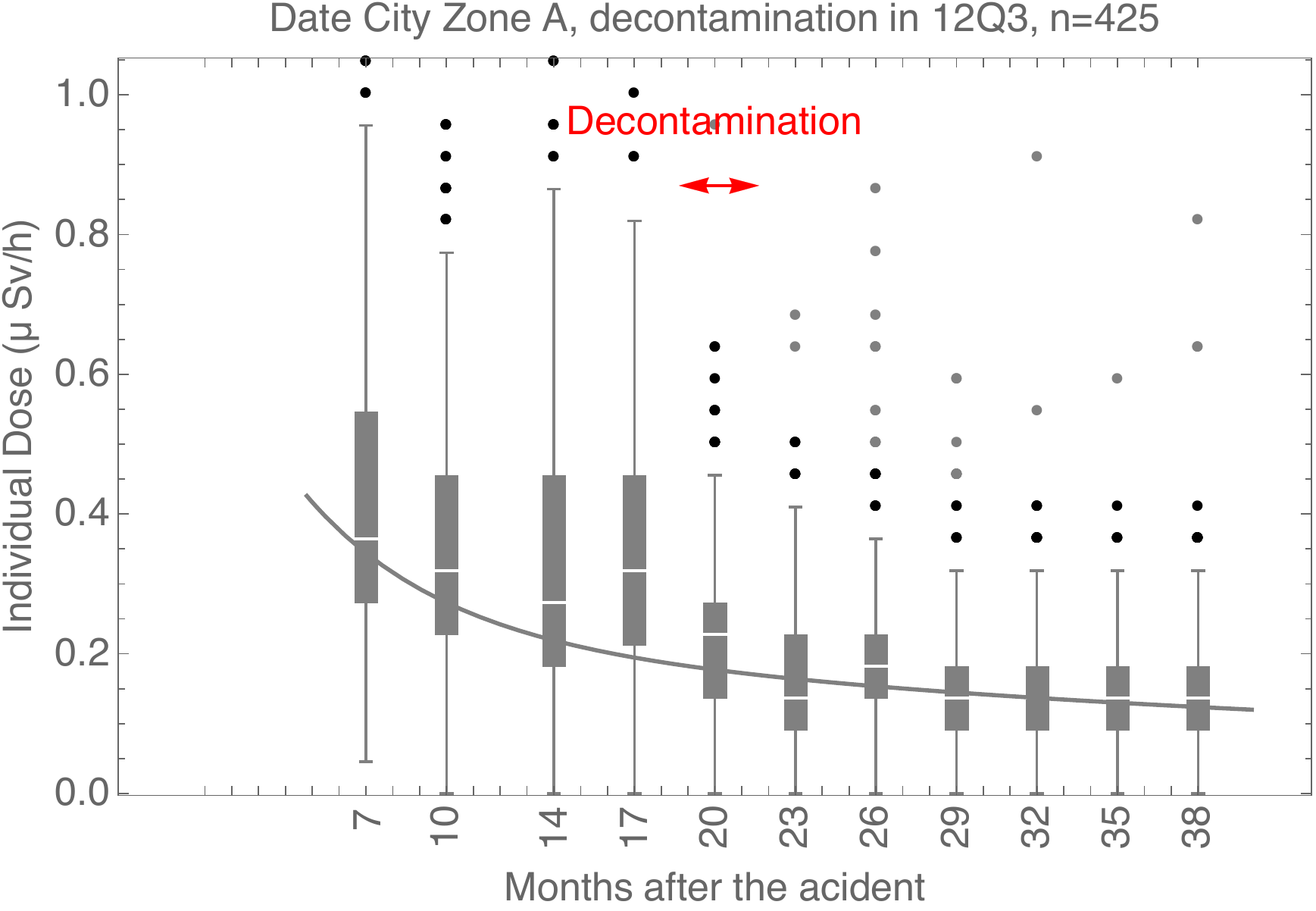}
\caption{\label{fig6}\footnotesize A box-and-whisker plot of the distribution of individual doses of 425 people who lived in zone A, whose houses were decontaminated during Q3 of  2012 (indicated by an red arrow); the measured value of the glass badge for 3 months is converted to the dose rate per hour. The superimposed curve, $\dot H_p (t)$ was calculated for zone A using the median grid dose $\dot H_{10}^{i\, A} (0.65)$ and the coefficient $c^A$, and hence  contains no adjustable parameters. As shown, the median values of individual doses are in good agreement with the reduction curve, except for that of 2012 Q2 (17 months); the reason for this deviation was not clear from the data used for the present analysis.\\}

\includegraphics[width=.5\textwidth]{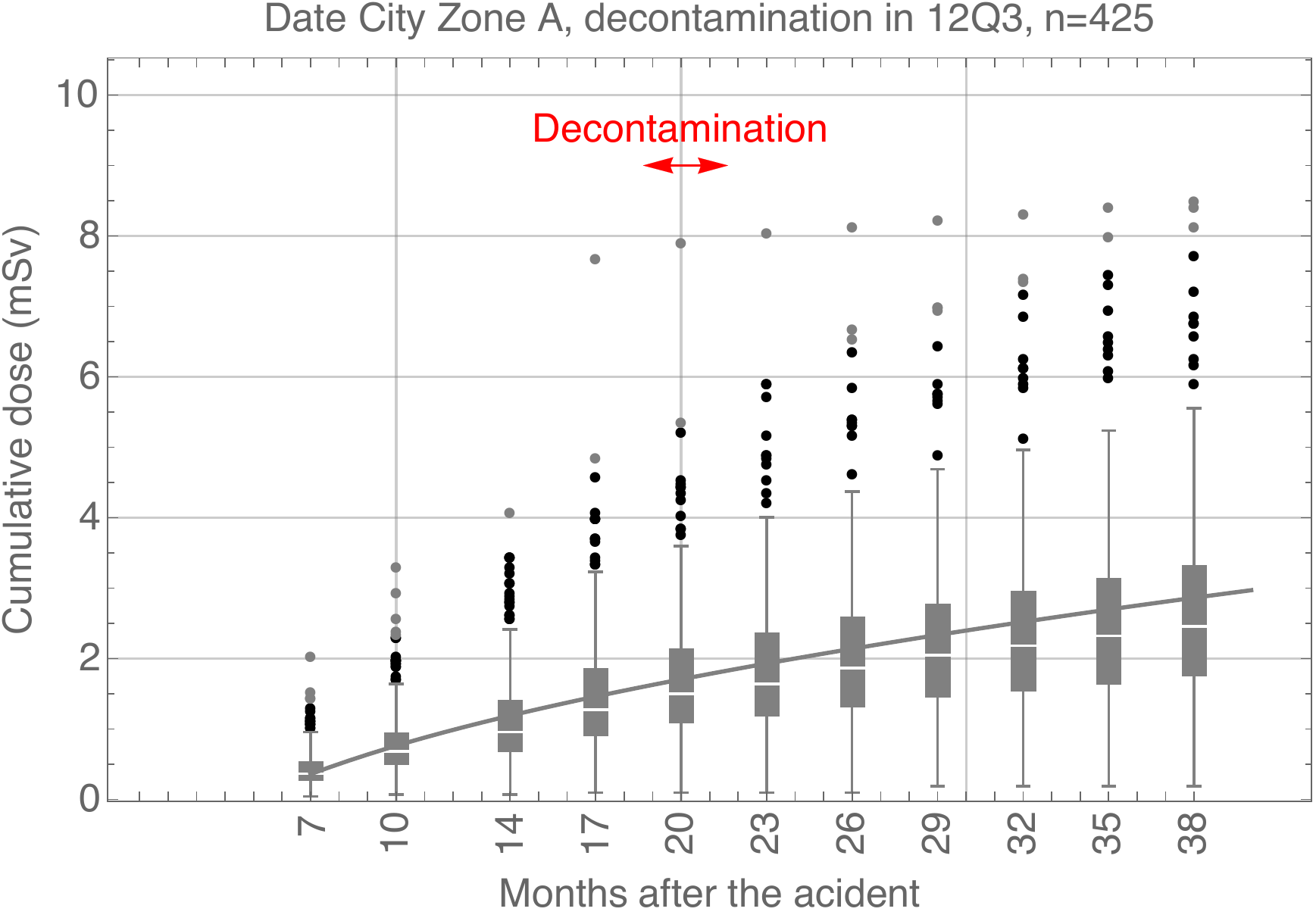}
\caption{\footnotesize The glass-badge data and the reduction curve shown in Fig.~\ref{fig6} are converted to  cumulative dose distributions and $H_P^A (t)$, respectively. The $H_P^A (t)$ curve is the same as was used in Fig.~\ref{fig5} 1). In this cumulative distribution, the deviation observed for 2012 Q2 in Fig.~\ref{fig5} is no longer dominant.}
\end{figure}
\clearpage
\section{DISCUSSION}

The current study has a number of unique characteristics. First, no other cases of continuous population-scale mass monitoring of the individual external doses of residents living in long-term contaminated areas have been recorded, meaning that no comparable individual dose data is available. 
After the Chernobyl accident, for example,  external doses were generally estimated based on the ambient dose rates~\cite{iaea}, rather than on individual dose monitoring. Therefore, the lifetime cumulative doses of residents, important information for those living in contaminated areas, were not derived from measured individual doses. 

The scale of the decontamination carried out in Fukushima Prefecture after the FDNPP accident is also unprecedented, meaning that comparable quantitative data regarding the effectiveness of decontamination in reducing individual doses on a population scale is unavailable.
In Japan after the FDNPP accident, the dose reduction  due to decontamination has until now generally been evaluated by comparing the ambient dose rates before and after decontamination, measured using survey meters at various outdoor points on the premises. As such, evaluation of the effect of decontamination on the individual doses of the residents has until now not been implemented either retrospectively or prospectively.

Fortunately, however, except for the initial 4 months after the FDNPP accident, individual doses have been continuously recorded in many municipalities in Fukushima Prefecture. In most cities, towns and villages, these individual measurements targeted limited groups of subjects and time periods. Date City, which began measuring personal doses for schoolchildren and residents living in relatively high-dose areas from August 2011 and has continued to measure tens of thousands of citizens uninterruptedly, is an important exception. The availability of appropriate population-scale individual dose data, the scale of the decontamination that was performed in the areas where the subjects reside, and the availability of consistent aerial survey data from before and after the decontamination period, make the current study possible. 

We reported in  the first installment of this series~\cite{miyazaki2016} that the correlation between the individual doses measured with  glass badges in Date City and the ambient dose rates derived from the  airborne monitoring of  residences did not change over time. Based on that conclusion, in this report, we estimated the cumulative doses over  a lifetime by extrapolating the individual doses of the Date City residents who continuously  held a glass badge, by making use of Eq.~(2). As a result, the median of the additional lifetime doses  from external exposure pathways for residents continuing to live in Zone A, the most contaminated area in Date City (and who received the highest doses in Fukushima Prefecture after the FDNPP accident), was estimated to be 18 mSv. The ICRP adopts an allowable dose band of 1 to 20 mSv per year for individuals in existing exposure situations~\cite{icrp103}.  The present study
indicates that the median additional lifetime (i.e., 70\,y) dose by the FNDPP accident of Date City residents will not exceed 20 mSv.

In the Chernobyl accident, it was estimated that the residents in Russia, Belarus and Ukraine  received about a quarter of their total lifetime exposure (up to 70 years after the accident) during the first year following the accident~\cite{iaea}. Similarly, the 1st-year external cumulative dose of Date City residents was estimated to be $\sim 1/6$ of their total lifetime exposure.

In August 2011, the Cabinet Office of Japan established a basic policy of decontamination, 
aiming at reducing the external dose rate received by the general public  by 50\%  (including physical decay and weathering effects) in the two years from the end of August 2011 to the end of August 2013. The report published in March 2015 by the Ministry of the Environment (MoE: responsible for decontamination)~\cite{moereport}, showed that  the additional external dose rate as estimated from  survey meter measurements decreased by approximately 62\% in those two years  in the area decontaminated according to the guidelines, to which the decontamination was said to contribute about 22\%, and physical decay etc. contributed 40\%. In Date City, the reduction of the grid dose (airborne monitoring survey) in the city between November 2011 and October 2013 was 60\%, and the reduction of the individual dose (glass badge) in Zone A during the same period was 62\%, similar to the result shown in the MoE report.

It is possible, however, to explain the 60\% reduction of the ambient dose rate in Date City by decay and weathering alone.  In addition, comparison between the individual- and  ambient-dose rates in Date City in the present study failed to confirm a 22\% reduction of the individual doses due to decontamination, for the residents in Zone A, where the decontamination was carried out according to the government guidelines. The authors believe that reasons why no decontamination effect could  be seen in Date City, unlike the results presented in the MoE report, include: 
\begin{enumerate}
\item The ratio of the areas actually decontaminated  to the area covered by the airborne monitoring is small. Thus, decontamination has little effect in reducing the ``grid'' dose.
\item The  residents do not necessarily spend all their time only in the decontaminated residential areas, and
\item the  measurements included in the MoE report were done at outdoor points on the premises, not reflecting the effect of decontamination on indoor dose rates.
\end{enumerate}
The information available for the present study, however, does not make it possible to evaluate the behavior patterns of  individual residents or the retention status of each dosimeter, so it is difficult to further discuss the reasons why the effect of decontamination is not reflected in the individual doses.
No comparable data is yet available to help determine whether or not this would be the case in other areas.  But because residents of Zone A in Date City received some of the highest external doses to the public, and decontamination was performed relatively early there, meaning its effects would likely be most evident, the authors believe that their findings may possibly prove to be representative.

\section{CONCLUSIONS}

In the present paper, we have shown that the ambient-dose-rate reduction function obtained from the airborne monitoring data can be applied uniformly to all zones in Date City. Regardless of the magnitude of the ambient dose rate, the difference in decontamination method, or whether or not decontamination was carried out, the reduction function was the same throughout Date City. Based on these results, the lifetime additional doses can be realistically estimated from the measured  individual doses, extrapolated by means of the  ambient-dose-rate reduction function. For establishing this method, which  combines the individual dose and the ambient dose, the existence of large-scale and long-term glass-badge data of Date City was essential.  Although the proportionality coefficient $c=0.15$ may not be universal, the authors believe that   the present method will make it possible in the future to reliably estimate the lifetime doses of residents living in contaminated areas based on regular airborne monitoring surveys, supplemented by smaller-scale individual dose monitoring surveys for the purpose of determining the ambient-dose-to-individual-dose coefficient $c$.

\begin{acknowledgments}

The authors are grateful to Chiyoda Technol Corporation, and Dr. J. Tada of Radiation Safety Forum for valuable discussions.

\end{acknowledgments}

\end{document}